\documentclass[preprint2]{aastex}

\begin{document}

\title{A Search for Neutrinos from the Solar {\it hep} Reaction and the \\
Diffuse Supernova Neutrino Background with the Sudbury Neutrino Observatory}

\shorttitle{A search for hep and DSNB neutrinos with SNO}
\shortauthors{B.~Aharmim {\em et al.}, The SNO Collaboration}

%Author list for hep paper: 6.5  July 7, 2006
\author{
B.~Aharmim\altaffilmark{\ref{lu}},
S.N.~Ahmed\altaffilmark{\ref{queens}},
A.E.~Anthony\altaffilmark{\ref{uta}},
E.W.~Beier\altaffilmark{\ref{penn}},
A.~Bellerive\altaffilmark{\ref{carleton}},
M.~Bergevin\altaffilmark{\ref{lbnl},\ref{uog}},
S.D.~Biller\altaffilmark{\ref{oxford}},
M.G.~Boulay\altaffilmark{\ref{queens},\ref{lanl}},
Y.D.~Chan\altaffilmark{\ref{lbnl}},
M.~Chen\altaffilmark{\ref{queens}},
X.~Chen\altaffilmark{\ref{lbnl},\ref{pny}},
B.T.~Cleveland\altaffilmark{\ref{oxford}},
G.A.~Cox\altaffilmark{\ref{uw}},
C.A.~Currat\altaffilmark{\ref{lbnl}},
X.~Dai\altaffilmark{\ref{queens},\ref{oxford}},
F.~Dalnoki-Veress\altaffilmark{\ref{carleton},\ref{mpi}},
H.~Deng\altaffilmark{\ref{penn}},
J.~Detwiler\altaffilmark{\ref{uw}},
M.~DiMarco\altaffilmark{\ref{queens}},
P.J.~Doe\altaffilmark{\ref{uw}},
G.~Doucas\altaffilmark{\ref{oxford}},
P.-L.~Drouin\altaffilmark{\ref{carleton}},
F.A.~Duncan\altaffilmark{\ref{snoi},\ref{queens}},
M.~Dunford\altaffilmark{\ref{penn}},
J.A.~Dunmore\altaffilmark{\ref{oxford},\ref{ucia}},
E.D.~Earle\altaffilmark{\ref{queens}},
H.C.~Evans\altaffilmark{\ref{queens}},
G.T.~Ewan\altaffilmark{\ref{queens}},
J.~Farine\altaffilmark{\ref{lu}},
H.~Fergani\altaffilmark{\ref{oxford}},
F.~Fleurot\altaffilmark{\ref{lu}},
R.J.~Ford\altaffilmark{\ref{snoi},\ref{queens}},
J.A.~Formaggio\altaffilmark{\ref{mitt},\ref{uw}},
N.~Gagnon\altaffilmark{\ref{uw},\ref{lanl},\ref{lbnl},\ref{oxford}},
J.TM.~Goon\altaffilmark{\ref{lsu}},
K.~Graham\altaffilmark{\ref{queens}},
E. ~Guillian\altaffilmark{\ref{queens}},
R.L.~Hahn\altaffilmark{\ref{bnl}},
A.L.~Hallin\altaffilmark{\ref{queens}},
E.D.~Hallman\altaffilmark{\ref{lu}},
P.J.~Harvey\altaffilmark{\ref{queens}},
R.~Hazama\altaffilmark{\ref{uw},\ref{rcnp}},
K.M.~Heeger\altaffilmark{\ref{uw},\ref{lbnl}},
W.J.~Heintzelman\altaffilmark{\ref{penn}},
J.~Heise\altaffilmark{\ref{snoi}},
R.L.~Helmer\altaffilmark{\ref{triumf},\ref{ubc}},
R.J.~Hemingway\altaffilmark{\ref{carleton}},
R.~Henning\altaffilmark{\ref{lbnl}},
A.~Hime\altaffilmark{\ref{lanl}},
C.~Howard\altaffilmark{\ref{queens}},
M.A.~Howe\altaffilmark{\ref{uw}},
M.~Huang\altaffilmark{\ref{uta}},
P.~Jagam\altaffilmark{\ref{uog}},
N.A.~Jelley\altaffilmark{\ref{oxford}},
J.R.~Klein\altaffilmark{\ref{uta},\ref{penn}},
L.L.~Kormos\altaffilmark{\ref{queens}},
M.~Kos\altaffilmark{\ref{queens}},
A.~Kr\"{u}ger\altaffilmark{\ref{lu}},
C.~Kraus\altaffilmark{\ref{queens}},
C.B.~Krauss\altaffilmark{\ref{queens}},
T.~Kutter\altaffilmark{\ref{lsu}},
C.C.M.~Kyba\altaffilmark{\ref{penn}},
H.~Labranche\altaffilmark{\ref{uog}},
R.~Lange\altaffilmark{\ref{bnl}},
J.~Law\altaffilmark{\ref{uog}},
I.T.~Lawson\altaffilmark{\ref{snoi},\ref{uog}},
K.T.~Lesko\altaffilmark{\ref{lbnl}},
J.R.~Leslie\altaffilmark{\ref{queens}},
J.C.~Loach\altaffilmark{\ref{oxford}},
S.~Luoma\altaffilmark{\ref{lu}},
R.~MacLellan\altaffilmark{\ref{queens}},
S.~Majerus\altaffilmark{\ref{oxford}},
H.B.~Mak\altaffilmark{\ref{queens}},
J.~Maneira\altaffilmark{\ref{lifep}},
A.D.~Marino\altaffilmark{\ref{lbnl},\ref{fnal}},
R.~Martin\altaffilmark{\ref{queens}},
N.~McCauley\altaffilmark{\ref{penn},\ref{liverpool}},
A.B.~McDonald\altaffilmark{\ref{queens}},
S.~McGee\altaffilmark{\ref{uw}},
C.~Mifflin\altaffilmark{\ref{carleton}},
K.K.S.~Miknaitis\altaffilmark{\ref{uw},\ref{uc}},
M.L.~Miller\altaffilmark{\ref{mitt}},
B.~Monreal\altaffilmark{\ref{mitt}},
B.G.~Nickel\altaffilmark{\ref{uog}},
A.J.~Noble\altaffilmark{\ref{queens}},
E.B.~Norman\altaffilmark{\ref{lbnl},\ref{llnl}},
N.S.~Oblath\altaffilmark{\ref{uw}},
C.E.~Okada\altaffilmark{\ref{lbnl},\ref{pnv}},
H.M.~O'Keeffe\altaffilmark{\ref{oxford}},
G.D.~Orebi~Gann\altaffilmark{\ref{oxford}},
S.M.~Oser\altaffilmark{\ref{ubc}},
R.~Ott\altaffilmark{\ref{mitt}},
S.J.M.~Peeters\altaffilmark{\ref{oxford}},
A.W.P.~Poon\altaffilmark{\ref{lbnl}},
G.~Prior\altaffilmark{\ref{lbnl}},
K.~Rielage\altaffilmark{\ref{lanl},\ref{uw}},
B.C.~Robertson\altaffilmark{\ref{queens}},
R.G.H.~Robertson\altaffilmark{\ref{uw}},
E.~Rollin\altaffilmark{\ref{carleton}},
M.H.~Schwendener\altaffilmark{\ref{lu}},
J.A.~Secrest\altaffilmark{\ref{penn}},
S.R.~Seibert\altaffilmark{\ref{uta}},
O.~Simard\altaffilmark{\ref{carleton}},
C.J.~Sims\altaffilmark{\ref{oxford}},
D.~Sinclair\altaffilmark{\ref{carleton},\ref{triumf}},
P.~Skensved\altaffilmark{\ref{queens}},
R.G.~Stokstad\altaffilmark{\ref{lbnl}},
L.C.~Stonehill\altaffilmark{\ref{lanl},\ref{uw}},
G.~Te\v{s}i\'{c}\altaffilmark{\ref{carleton}},
N.~Tolich\altaffilmark{\ref{lbnl}},
T.~Tsui\altaffilmark{\ref{ubc}},
R.~Van~Berg\altaffilmark{\ref{penn}},
R.G.~Van~de~Water\altaffilmark{\ref{lanl}},
B.A.~VanDevender\altaffilmark{\ref{uw}},
C.J.~Virtue\altaffilmark{\ref{lu}},
T.J.~Walker\altaffilmark{\ref{mitt}},
B.L.~Wall\altaffilmark{\ref{uw}},
D.~Waller\altaffilmark{\ref{carleton}},
H.~Wan~Chan~Tseung\altaffilmark{\ref{oxford}},
D.L.~Wark\altaffilmark{\ref{ralimp},\ref{imp}},
J.~Wendland\altaffilmark{\ref{ubc}},
N.~West\altaffilmark{\ref{oxford}},
J.F.~Wilkerson\altaffilmark{\ref{uw}},
J.R.~Wilson\altaffilmark{\ref{oxford},\ref{suss}},
J.M.~Wouters\altaffilmark{\ref{lanl}},
A.~Wright\altaffilmark{\ref{queens}},
M.~Yeh\altaffilmark{\ref{bnl}},
F.~Zhang\altaffilmark{\ref{carleton}},
K.~Zuber\altaffilmark{\ref{oxford},\ref{suss}}
}

\altaffiltext{1}{ 
 \label{ubc} Department of Physics and Astronomy, University of British Columbia, Vancouver, BC V6T 1Z1, Canada  
 }

 \altaffiltext{2}{ 
 \label{bnl} Chemistry Department, Brookhaven National Laboratory, Upton, NY 11973-5000  
 }

 \altaffiltext{3}{ 
 \label{carleton} Ottawa-Carleton Institute for Physics, Department of Physics, Carleton University, Ottawa, Ontario K1S 5B6, Canada  
 }

 \altaffiltext{4}{ 
 \label{uog} Physics Department, University of Guelph, Guelph, Ontario N1G 2W1, Canada  
 }

 \altaffiltext{5}{ 
 \label{lu} Department of Physics and Astronomy, Laurentian University, Sudbury, Ontario P3E 2C6, Canada  
 }

 \altaffiltext{6}{ 
 \label{lbnl} Institute for Nuclear and Particle Astrophysics and Nuclear Science Division, Lawrence Berkeley National Laboratory, Berkeley, CA 94720  
 }

 \altaffiltext{7}{ 
 \label{lanl} Los Alamos National Laboratory, Los Alamos, NM 87545  
 }

 \altaffiltext{8}{ 
 \label{lifep} Laborat\'{orio de Instrumenta\c{c}\~{a}o e F\'{i}sica Experimenttal de Part\'{i}culas, Lisboa, Portugal}  
 }

 \altaffiltext{9}{ 
 \label{lsu} Department of Physics and Astronomy, Louisiana State University, Baton Rouge, LA 70803  
 }

 \altaffiltext{10}{ 
 \label{mitt} Laboratory for Nuclear Science, Massachusetts Institute of Technology, Cambridge, MA 02139  
 }

 \altaffiltext{11}{ 
 \label{oxford} Department of Physics, University of Oxford, Denys Wilkinson Building, Keble Road, Oxford OX1 3RH, UK  
 }

 \altaffiltext{12}{ 
 \label{penn} Department of Physics and Astronomy, University of Pennsylvania, Philadelphia, PA 19104-6396  
 }

 \altaffiltext{13}{ 
 \label{queens} Department of Physics, Queen's University, Kingston, Ontario K7L 3N6, Canada  
 }
\altaffiltext{14}{ 
 \label{ralimp} Rutherford Appleton Laboratory, Chilton, Didcot OX11 0QX, UK  
 }

 \altaffiltext{15}{ 
 \label{uw} Center for Experimental Nuclear Physics and Astrophysics, and Department of Physics, University of Washington, Seattle, WA 98195  
 }

 \altaffiltext{16}{ 
 \label{snoi} SNOLAB, P.O. Box 146, Lively, ON P3Y 1M3, Canada  
 }

 \altaffiltext{17}{ 
 \label{uta} Department of Physics, University of Texas at Austin, Austin, TX 78712-0264  
 }

 \altaffiltext{18}{ 
 \label{triumf} TRIUMF, 4004 Wesbrook Mall, Vancouver, BC V6T 2A3, Canada  
 }

 \altaffiltext{19}{ 
 \label{pny} Present address: Goldman Sachs, 85 Broad Street, New York, NY 
 }

 \altaffiltext{20}{ 
 \label{mpi} Present address:  Physics Department, Princeton University, Princeton, NJ 08544-0708 
 }

 \altaffiltext{21}{ 
 \label{ucia} Present address: Department of Physics, University of California, Irvine, CA  
 }

 \altaffiltext{22}{ 
 \label{rcnp} Present address: Graduate School of Engineering, Hiroshima University, Hiroshima, Japan
 }

 \altaffiltext{23}{ 
 \label{fnal} Present address: Fermilab, Batavia, IL  
 }

 \altaffiltext{24}{ 
 \label{liverpool} Present address: Department of Physics, University of Liverpool, Liverpool, L69 7ZE, UK  
 }

 \altaffiltext{25}{ 
 \label{uc} Present address: Department of Physics, University of Chicago, Chicago, IL  
 }

 \altaffiltext{26}{ 
 \label{llnl} Present address: Lawrence Livermore National Laboratory, Livermore, CA  
 }

 \altaffiltext{27}{ 
 \label{pnv} Present address: Remote Sensing Lab, P.O. Box 98521, Las Vegas, NV 89193 
 }

 \altaffiltext{28}{ 
 \label{suss} Present address: Department of Physics and Astronomy, University of Sussex, Brighton BN1 9QH, UK  
 }
 \altaffiltext{29}{ 
 \label{imp} Alternate address: Imperial College, London SW7 2AZ, UK }

\begin{abstract}

A search has been made for neutrinos from the {\it hep} reaction in
the Sun and from the diffuse supernova neutrino background (DSNB)
using data collected during the first operational phase of the Sudbury
Neutrino Observatory, with an exposure of 0.65 kilotonne-years. For the
{\it hep} neutrino search, two events are observed in the effective
electron energy range of 14.3~MeV $<\mathrm{T_{\mathrm{eff}}}<$ 20~MeV
where 3.1 background events are expected. After accounting for
neutrino oscillations, an upper limit of
$2.3\times10^4$~cm$^{-2}$s$^{-1}$ at the 90$\%$ confidence level is
inferred on the integral total flux of {\it hep} neutrinos. For DSNB neutrinos,
no events are observed in the effective electron energy range of
21~MeV $<\mathrm{T_{\mathrm{eff}}}<$ 35~MeV and, consequently, an
upper limit on the $\nu_e$ component of the DSNB flux in the neutrino
energy range of 22.9~MeV $<E_{\nu}<$ 36.9~MeV of 70~cm$^{-2}$s$^{-1}$
is inferred at the 90$\%$ confidence level. This is an improvement by
a factor of 6.5 on the previous best upper limit on the {\it hep}
neutrino flux and by two orders of magnitude on the previous upper
limit on the $\nu_e$ component of the DSNB flux.

\end{abstract}

\keywords{neutrinos, Sun: general, supernovae: general}

%%%%%%%%%%%%%%%%%%%%%%%%%%%%%%%%%%%%%%%%%%%%%%%%%%%%%%%%%%%%%%%%%%%%%%%%%%
%                           Introduction                                 %
%%%%%%%%%%%%%%%%%%%%%%%%%%%%%%%%%%%%%%%%%%%%%%%%%%%%%%%%%%%%%%%%%%%%%%%%%%

\section{Introduction}

The Sudbury Neutrino Observatory (SNO) is a real-time, heavy water
Cherenkov detector located in the Inco, Ltd.~Creighton nickel mine
near Sudbury, Ontario, Canada at a depth of 6010 m water
equivalent~(\cite{sno_nim}). SNO detects electrons and neutrons from,
respectively, charged-current (CC) and neutral-current (NC)
interactions of neutrinos on deuterons, as well as neutrino-electron
elastic scattering (ES) interactions, in one kilotonne of D$_2$O
contained in a 12~m diameter acrylic vessel (AV). These interactions
are observed via Cherenkov light detected by 9456 photomultiplier
tubes (PMTs) mounted on a 17.8~m diameter support structure.  By
comparing the observed rates of these interactions, SNO has
demonstrated that a substantial fraction of the $^8$B electron
neutrinos produced in the Sun transform into other active neutrino
flavors~(\cite{cc_prl,d2o_prl,dn_prl,salt_prl,nsp}). These results are
consistent with the predictions of neutrino
oscillations~(\cite{mnsp1,mnsp2,msw1,msw2}).

The Sun generates energy by nuclear fusion; protons combine to form
helium in reactions that release neutrinos. The primary solar fusion
process is a series of reactions known as the pp chain. Five reactions
in the pp chain produce neutrinos; the highest energy neutrinos are
those from the {\it hep} reaction: $^3$He + p $\rightarrow$ $^4$He +
e$^+$ + $\nu_e$. The endpoint of the {\it hep} neutrino spectrum is
18.77~MeV and lies above that of the $^8$B spectrum, which is
approximately 15~MeV. The flux of {\it hep} neutrinos ({\it e.g.} 
\cite{bk98}, \cite{bp2004}) is currently predicted to be
$(7.97\pm1.24)\times10^3$~cm$^{-2}$s$^{-1}$~(\cite{solar_model}),
which is small compared to the fluxes from the other
neutrino-producing reactions in the pp chain, including the $^8$B flux
which has been measured to be
$(4.95\pm0.42)\times10^6$~cm$^{-2}$s$^{-1}$~(\cite{nsp}).  The
dominant contribution to the uncertainty in the {\it hep} neutrino
flux prediction is $15.1\%$ from the calculation of the nuclear matrix
elements~(\cite{hep_nucl_me}).  The previous best upper limit on the
{\it hep} neutrino flux is $7.3\times10^4$~cm$^{-2}$s$^{-1}$ at the
$90\%$ confidence level (CL), based on measurements with the
Super-Kamiokande detector~(\cite{sk-long}). After accounting for
neutrino oscillations, this limit can be interpreted as an upper bound
on the total flux of {\it hep} neutrinos of
$1.5\times10^5$~cm$^{-2}$s$^{-1}$. Currently, only one reaction
($^8$B) from the pp chain has been uniquely observed and measured
experimentally. An observation of {\it hep} neutrinos would give
further confirmation of the pp chain as the primary solar energy
generation mechanism and would allow further tests of the solar model.

Neutrinos produced in core-collapse supernovae also contribute to the
energy region above the $^8$B endpoint. The current generation of
neutrino detectors can detect the transient signal from a supernova in
the Milky Way, but the expected signal from a supernova in a more
distant galaxy is fewer than one event. Neutrinos from all
extragalactic supernovae since the beginning of the formation of stars
in the universe constitute the diffuse supernova neutrino background
(DSNB), which may be detectable. Model predictions range from 0.19 to
1.49~cm$^{-2}$s$^{-1}$ for the $\nu_e$ component of the DSNB flux in
the neutrino energy range 22.9~MeV $<E_{\nu}<$ 36.9~MeV
(\cite{beacom-dsnb},~\cite{ando-dsnb}). The best upper limit on the
$\overline{\nu}_e$ component of the DSNB flux is 1.2~cm$^{-2}$s$^{-1}$
at the 90$\%$ CL for $E_{\bar{\nu}}> 19.3$~MeV, based on measurements
with the Super-Kamiokande detector~(\cite{sk-dsnb}). While an
indirect limit on the $\nu_e$ component of the DSNB flux can be
inferred from this (\cite{lunardini}), the previous best direct
upper limit is $6.8\times10^{3}$~cm$^{-2}$s$^{-1}$ for neutrino
energies 25~MeV $<E_{\nu}<$ 50~MeV, based on measurements with the
Mont Blanc liquid scintillator detector~(\cite{mb-dsnb}).

A search for {\it hep} and DSNB neutrinos has been performed by
counting the numbers of events in predefined energy intervals (signal
boxes) above the $^8$B endpoint. The most sensitive signal boxes for
this analysis were selected by evaluating the predicted signal and
background levels before examining the data. Given the predicted
signal and background levels in the signal boxes, limits on the flux
of {\it hep} and DSNB neutrinos are set using a modified
Feldman-Cousins technique. The following sections describe the data
set, detector response, determination of the backgrounds, analysis
procedures and the limits obtained for the {\it hep} and DSNB neutrino
fluxes.

%%%%%%%%%%%%%%%%%%%%%%%%%%%%%%%%%%%%%%%%%%%%%%%%%%%%%%%%%%%%%%%%%%%%%%%%%%%%
%                             Data Set                                     %
%%%%%%%%%%%%%%%%%%%%%%%%%%%%%%%%%%%%%%%%%%%%%%%%%%%%%%%%%%%%%%%%%%%%%%%%%%%%
\section{The Data Set}
\label{sec:data_sets}

The data included in these analyses were collected during the initial
phase of SNO operation, during which the detector contained pure
D$_2$O. The data were collected from 1999 November 2 until 2001 May 28
and comprise 306.4 live days corresponding to an exposure of 0.65
kilotonne-years~(\cite{d2o_prl}).

Since results from this phase were last published, numerous
improvements have been made to the analysis tools, many of which were
used in the analysis of data from phase two~(\cite{nsp}), for which
two tonnes of salt were dissolved in the heavy water. Further
improvements were applied in this analysis, the most significant of
which was improved estimation of the effective electron kinetic
energies ($\mathrm{T_{\mathrm{eff}}}$) of the events, based on the
optical paths to each operational PMT. Other enhancements include
improved handling of false hits due to cross-talk between electronics
channels and an improved accounting of working PMTs using both
neutrino and calibration data to track bad channels. However, the
vertex reconstruction algorithm was the same as that used in previous
phase one analyses, in which events were reconstructed under the
assumption that they are due to single electrons.  This is more suited
for the reconstruction of {\it hep} and DSNB events than the algorithm
used in phase two. After the application of the new analysis tools,
events inside the kinetic energy window of 12~MeV
$<\mathrm{T_{\mathrm{eff}}}<$ 35~MeV were not examined until the {\it
hep} and DSNB signal boxes had been selected.

In addition to the event selection discussed in~\cite{nsp}, which
includes a selection that removes Michel electrons with visible
precursors, selection criteria were applied to remove backgrounds from
atmospheric neutrino interactions. As the {\it hep} and DSNB signals
are expected to be single electron events, these backgrounds can be
reduced significantly by removing events which correlate in time with
neutrons, electrons or $\gamma$-rays.  Consequently, any candidate
event that appeared within 250~ms of another with
$\mathrm{T_{\mathrm{eff}}}> 4$~MeV and a reconstructed vertex inside
the AV was removed. In addition, two Kolmogorov-Smirnov (KS) tests
were applied: one to test the azimuthal symmetry of the PMT hits about
the reconstructed event direction; the other to test the compatibility
of the angular distribution of PMT hits with that expected from a
single electron. In the signal boxes, the selections on PMT hit
isotropy and the prompt light fraction were further tightened with
respect to previous SNO analyses (\cite{nsp}). This was possible in
this analysis due to the higher energies of the candidate events. The
combined event selection reduced the expected number of atmospheric
neutrino events in the {\it hep} signal box by a factor of 29 and by a
factor of 77 in the DSNB signal box. The signal acceptance of the
combined event selection is $(96.6\pm0.7)\%$ for {\it hep} and
$(94.0\pm1.5)\%$ for DSNB events, measured using calibration source
data and simulation.

%%%%%%%%%%%%%%%%%%%%%%%%%%%%%%%%%%%%%%%%%%%%%%%%%%%%%%%%%%%%%%%%%%%%%%%%%%
%                           Detector Response                            %
%%%%%%%%%%%%%%%%%%%%%%%%%%%%%%%%%%%%%%%%%%%%%%%%%%%%%%%%%%%%%%%%%%%%%%%%%%

\section{Detector Response}
\label{sec:response}

\begin{figure}
\begin{center}
\includegraphics[width=3.3in]{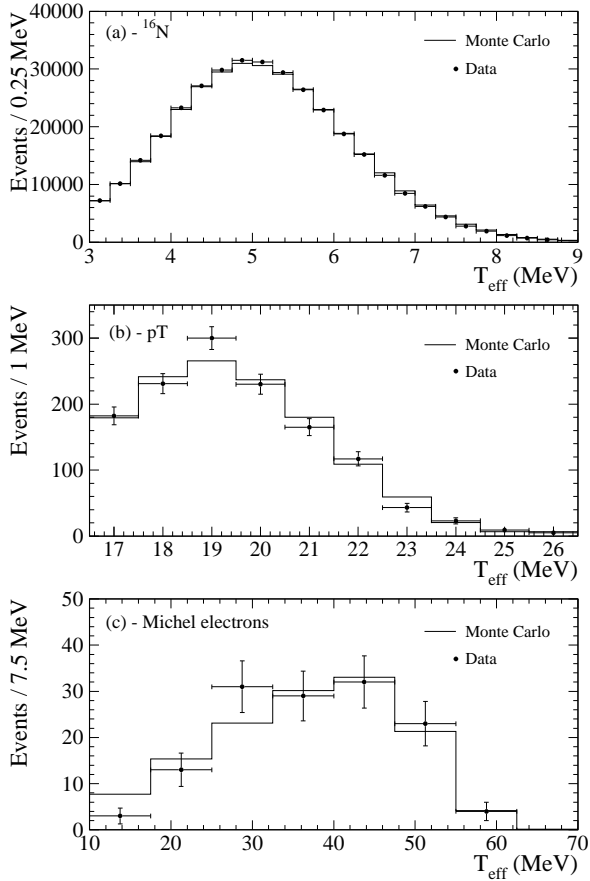}
\caption{\label{fig:hep_ene_cal}
The effective electron kinetic energy spectra from data and Monte Carlo for 
(a) events from the $^{16}$N source, (b) events from the pT source and (c) 
Michel electrons. The data are shown in the energy regions free of source 
related backgrounds.}
\end{center}
\end{figure}

To understand the signals and backgrounds in this analysis, it is
important to measure the energy response and uncertainties in the
signal boxes. The energy response can be parameterized by a Gaussian
of resolution $\sigma_T = -0.154 + 0.390\sqrt{T_{e}} + 0.0336T_{e}$,
where $T_{e}$ is the true kinetic energy of the electron. In SNO
analyses, Monte Carlo simulation is used to estimate the response of
the detector to different particles. The propagation of electrons,
positrons and $\gamma$-rays is carried out using EGS4~(\cite{egs4}). The
uncertainties in the energy scale and resolution of the SNO detector
have typically been measured using 6.13~MeV $\gamma$-rays from a
$^{16}$N source~(\cite{sno-16n}).  At the higher energies more
characteristic of this analysis, Michel electrons from muon decays and
a pT ($^3$H(p,$\gamma$)$^4$He) source (\cite{sno-pt}), which produces
19.8~MeV $\gamma$-rays, were used to complement the $^{16}$N
measurements. Using simple event selection criteria, including one
based on the time between events, 135 Michel electrons were identified
in the data.  Potential deviations in energy scale and energy
resolution between data and simulations were assumed to be linear
functions of energy. These functions were fit with a maximum
likelihood technique using data from $^{16}$N and pT sources as
further constraints.  The results were used to refine the energy scale
and resolution estimates and to measure their uncertainties at the
analysis thresholds.  An energy scale uncertainty of $0.96\%$ and a
resolution uncertainty of $3.8\%$ were estimated at the {\it hep}
threshold of 14.3~MeV. At the DSNB threshold of 21~MeV, an energy
scale uncertainty of $1.06\%$ and a resolution uncertainty of $6.0\%$
were estimated. Correlations between these quantities were included in
the final analysis. Additional non-Gaussian tails to the resolution
function were also considered, but were found to be
insignificant. Data and Monte Carlo distributions of
$\mathrm{T_{\mathrm{eff}}}$ for $^{16}$N and pT calibration events and
for Michel electrons are shown in Figure~\ref{fig:hep_ene_cal}.

Event vertex and direction reconstruction were unchanged from the
analysis in~\cite{d2o_prl}. The position resolution at 15~MeV is
$({12.0}\pm{2.5})$~cm and the angular resolution is ($20.6\pm0.4)^{\rm
o}$. These were measured using a combination of $^{16}$N source data
and simulation. The same fiducial volume, defined by events
reconstructed within a distance of 550~cm from the center of the
detector, was selected. The uncertainty on the expected number of
events within the fiducial volume due to vertex accuracy was $2.9\%$.

%%%%%%%%%%%%%%%%%%%%%%%%%%%%%%%%%%%%%%%%%%%%%%%%%%%%%%%%%%%%%%%%%%%%%%%%%%
%                           Backgrounds                                  %
%%%%%%%%%%%%%%%%%%%%%%%%%%%%%%%%%%%%%%%%%%%%%%%%%%%%%%%%%%%%%%%%%%%%%%%%%%

\section{Backgrounds}
\label{sec:backgrounds}

\begin{figure}[tbp]
\begin{center}
\includegraphics[width=3.3in]{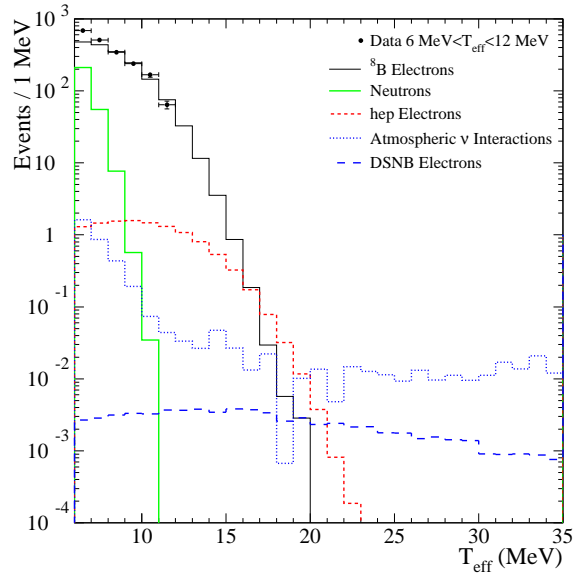}
\caption{\label{fig:hep_ene_pdfs}
The simulated effective electron kinetic energy spectra of the signals
and backgrounds of interest in the {\it hep} and DSNB analyses. Also
shown are the data in the range 6~MeV $<\mathrm{T_{\mathrm{eff}}}<$
12~MeV which are used to normalize the $^8$B electron and neutron
distributions. The atmospheric neutrino background is made up of a number of
different signals: neutrons at low energies; $\gamma$-rays from
nuclear de-excitations at intermediate energies; and Michel electrons
and CC interactions of atmospheric $\nu_e$ and $\bar{\nu_e}$ on
deuterons at higher energies. The DSNB model in this figure is the
$T=6$~MeV model from~\cite{beacom-dsnb}. The third class of
background, instrumental backgrounds, which is not shown in this figure is 
negligible.}
\end{center}
\end{figure}

Three distinct classes of background are considered: $^8$B neutrino
interactions, atmospheric neutrino interactions and instrumental
backgrounds.  Figure~\ref{fig:hep_ene_pdfs} shows the simulated energy
spectra of the signals and backgrounds, normalized to their expected
rates.

Electrons from $^8$B neutrino interactions are the dominant (97$\%$)
background for the {\it hep} analysis but are a negligible background
for DSNB. These events can reconstruct into the {\it hep} signal box
due to the finite energy resolution of the detector. The magnitude of
the $^8$B background depends on the details of the detector response
and is very sensitive to the energy scale and resolution at
threshold. In the CC interaction, by which SNO predominantly detects
the $^8$B and {\it hep} neutrinos, there is a strong correlation
between neutrino and electron energy. This, in addition to a cross
section that rises with the square of the energy rather than linearly,
provides a clearer distinction between the two neutrino spectra in
the region of the $^8$B endpoint than is possible with the ES
interaction.

The $^8$B background also depends on the details of the shape of the
detected electron spectrum. The $^8$B neutrino spectrum
from~\cite{b8spec,b8speclong} was assumed along with its quoted
uncertainties.  Neutrino oscillations were taken into account by
correcting and combining the electron spectra from CC and ES
interactions using the energy-dependent $\nu_e$ survival probability
from the joint solar neutrino and KamLAND (\cite{kamland}) oscillation
analysis presented in~\cite{nsp}. Additional spectral adjustments were
included to account for CC interactions on $^{18}$O, radiative
corrections to the CC deuteron cross section ~(\cite{nsa+,kmv}) and
the acceptance of the event selection. The expectation for the hep
signal is constructed in the same way using the neutrino spectrum from
\cite{hepspec} with corrections from \cite{hepspec_corr}.

After the determination of the $^8$B signal shape, its normalization
was determined using data at lower energies, where the {\it hep}
signal is insignificant. In an energy window of 6~MeV
$<\mathrm{T_{\mathrm{eff}}}<$ 12~MeV, 2006 events were observed. To
determine normalizations, these data were fit using a maximum
likelihood technique with probability density functions (PDFs) for the
$^8$B electrons (CC and ES signals) and neutrons (NC signal and
background). The distributions used in this fit were functions of
event energy and direction with respect to the Sun
($\cos\theta_{\odot}$). The results of this fit were then used to
estimate the $^8$B contribution inside the signal boxes.

Atmospheric neutrino interactions produce a second class of background events.
They are the dominant background in the
DSNB signal box and come from several sources:
\begin{itemize}
\item Electrons from low energy ($E_{\nu}<$ 100~MeV) charged-current 
$\nu_e$ and $\overline\nu_e$ interactions;
\item Michel electron events, where the precursor muons 
(and pions) are below the Cherenkov threshold and do not trigger the detector;
\item 15.1~MeV $\gamma$-rays from de-excitation of an excited state of 
$^{12}$C created via a nuclear cascade from neutrino interactions on $^{16}$O;
\item Mis-identified non-electron events.
\end{itemize}

For low energy atmospheric $\nu_e$ and $\overline\nu_e$, the flux
prediction from~\cite{lowe_atm} is used, which has an uncertainty of
$25\%$. Only charged current interactions on deuterons, with cross
sections from~\cite{nsa+} and ~\cite{kmv} are considered; the
contributions from other interaction types are not significant. The
interactions of these neutrinos constitute $14\%$ of the DSNB
background, but are insignificant in the {\it hep} signal box.

Events from the latter three sources are associated with atmospheric
neutrinos of higher energy ($E_{\nu}>$ 100~MeV). Monte Carlo simulations
were used to generate atmospheric neutrino interactions in the SNO
detector with statistics equivalent to 500 times the expected number
of events. For this purpose, the package NUANCE ~(\cite{nuance}) was
used, with the Bartol04 flux prediction for
Sudbury~(\cite{bartol04}). The flux uncertainty in the neutrino energy
range that contributes to the background is $10\%$. The events
generated by NUANCE were then propagated and fully simulated in the
SNO Monte Carlo, from which background predictions were obtained after
application of the event selection.

To assess uncertainties these events were divided into three
categories.  The first category, $\nu_\mu$ quasi-elastic (QE) CC
events, is the primary source of untagged Michel electrons and
originates from neutrinos in the energy range 150 to 250~MeV.  The
uncertainty on the cross section in this energy region is
$25\%$~(\cite{cc_atm_xs}). These Michel electrons comprise 80$\%$
of the DSNB background. For the second category,
15.1~MeV $\gamma$-ray events, there is no data in the literature on
production rates and, thus, a 100$\%$ uncertainty was assigned to the
production rate predicted by NUANCE, which uses the calculation of
\cite{15.1_gar}. These $\gamma$-rays constitute half of the atmospheric
background in the {\it hep} analysis, but due to the magnitude of
the $^8$B background constitute only 1.5$\%$ of the total {\it hep}
background. The final category comprises QE NC events and
interactions that produce pions, to which a cross section uncertainty
of $30\%$ is assigned~(\cite{nc_atm_xs}). 

There is an additional uncertainty applicable to the latter two
categories of atmospheric neutrino interactions. A comparison of
events from data and the simulation has shown that the simulation
underestimates the production of correlated neutrons. It is unclear if
this is due to errors in the prediction of primary neutron production
or in the transport of hadrons in the simulation. However, there is
good agreement between data and Monte Carlo for correlated electron
events. Events in the simulation are re-weighted in such a way that
the average neutron multiplicity is changed to better match the
data. This results in a change to the background rejection rate in the
simulation due to time correlated neutrons. This correction results in
an additional uncertainty of $7\%$ in the rate of atmospheric
background events inside the signal box that are not due to QE CC
interactions. After application of the event selection to remove
events with correlated neutrons, electrons and gamma-rays the
atmospheric background in these analyses is reduced by a factor of
two. 

To verify the predictions for the atmospheric neutrino background,
data outside the signal box in the energy range 35~MeV
$<\mathrm{T_{\mathrm{eff}}}<$ 55~MeV were examined. This energy range
was selected to be most sensitive to the main component of the
atmospheric neutrino background: the Michel
electrons. In this energy range, 0.28 Michel electrons and 0.05
electrons from low energy charged-current atmospheric neutrino
interactions are expected. One event was observed, consistent with the
predictions of the simulation. Inside this energy range, the effect of
the event selection on events correlated with neutrons, electrons or
$\gamma$-rays was also examined. Two such events were observed,
each consistent with being an otherwise untagged Michel electron
preceded by a $\gamma$-ray from the de-excitation of the nucleus
participating in the primary neutrino interaction. No events were
observed that are correlated with neutron or electron events. These
results are also consistent with the predictions of the simulation.

The final class of backgrounds is associated with instrumental effects
such as electronic pickup or static discharge from the PMTs. For these
events, an upper limit of 0.002 events is set in an energy range of
6~MeV $<\mathrm{T_{\mathrm{eff}}}<$ 35~MeV using the technique
described in~\cite{nsp}. This analysis is not sensitive to the
isotropic acrylic vessel background (IAVB) (see~\cite{nsp}). To
predict the number of IAVB events that pass the signal event
selection, the 13 IAVB events clearly identified in the data by simple
criteria are scaled via Monte Carlo simulation. A limit of
$7\times10^{-4}$ IAVB events in the energy range of 14~MeV
$<\mathrm{T_{\mathrm{eff}}}<$ 35~MeV is inferred at the $90\%$ CL.

%%%%%%%%%%%%%%%%%%%%%%%%%%%%%%%%%%%%%%%%%%%%%%%%%%%%%%%%%%%%%%%%%%%%%%%%%%
%                          Analysis and Results                          %
%%%%%%%%%%%%%%%%%%%%%%%%%%%%%%%%%%%%%%%%%%%%%%%%%%%%%%%%%%%%%%%%%%%%%%%%%%

\section{Analysis and results}
\label{sec:analysis}

The analysis was designed to construct confidence intervals on the
neutrino fluxes using a modified Feldman-Cousins
approach~(\cite{feldman-cousins,pole,hill}). Limits were also
calculated using a Bayesian approach (\cite{pdg}); very similar
results are obtained for the two techniques.  To determine
confidence limits the probability $p(N|S)$ of observing $N$ events,
given a signal flux $S$, is calculated taking statistical fluctuations
and all known systematic uncertainties into account. A Monte Carlo
technique is used to integrate over the estimated distributions of the
systematic uncertainties, including known correlations, by sampling
ensembles of shifted parameter values and propagating their effect on
the PDFs and extracted signal and background normalizations.  The
major uncertainties included in this procedure are shown in
Table~\ref{tab:sys}.

\begin{table} [h]
\tablewidth{7.7cm}
\begin{center}
\caption{\label{tab:sys} Major uncertainties included in the analyses.}
\begin{tabular}{ l l } 
\tableline
\tableline
Source of Uncertainty & Magnitude of effect \\
\tableline
Energy scale   & \\
~~~~$\mathrm{T_{\mathrm{eff}}}= 14.3$~MeV & $0.96\%$ \\
~~~~$\mathrm{T_{\mathrm{eff}}}= 21$~MeV & $1.06\%$ \\
Energy resolution & \\
~~~~$\mathrm{T_{\mathrm{eff}}}= 14.3$~MeV & $3.8\%$ \\
~~~~$\mathrm{T_{\mathrm{eff}}}= 21$~MeV & $6.0\%$ \\
Vertex accuracy & $2.9\%$ \\ 
Vertex resolution & $2.5$~cm \\
Angular resolution & $2\%$ \\
$^8$B $\nu_e$ spectrum & Taken from 1 \\ 
$\tan^2(\theta_{12})$,~$\Delta m^2_{12}$ & Contours from 2 \\
$\nu_{atm}$ flux &  \\
~~~~$E_{\nu}>$ 100~MeV& $10\%$ \\
~~~~$E_{\nu}<$ 100~MeV& $25\%$ \\
Cross sections & \\
~~~~CC deuteron & $1.2\%$ \\
~~~~$\nu_{atm}$ CC QE & $25\%$ \\
~~~~$\nu_{atm}$ other & $30\%$ \\
15.1~MeV $\gamma$-rays & $100\%$ \\
$\nu_{atm}$ n-multiplicity &  $7\%$ \\
\tableline
\end{tabular} 
\end{center}
\tablerefs{(1) \cite{b8speclong}; (2) \cite{nsp}}
\end{table}

The {\it hep} and DSNB analyses are very similar, except
that the definitions of signal and background are
modified. In the DSNB analysis, the {\it hep} distribution is scaled
using the standard solar model prediction, including its uncertainty,
and added to the background estimate.

\begin{figure}[tbp] 
\begin{center}
\includegraphics[width=3.3in]{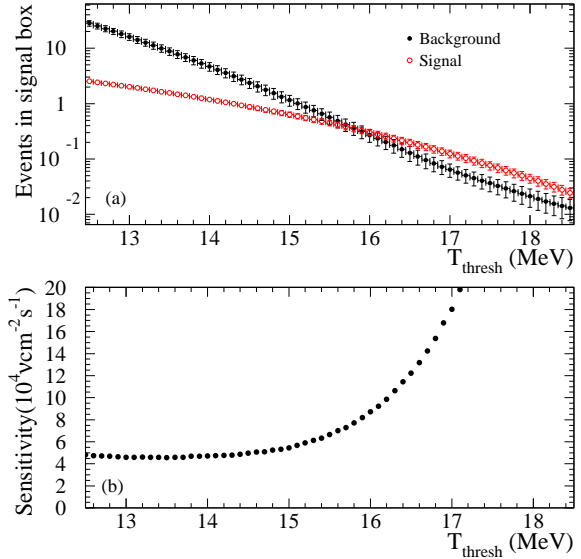}
\caption{\label{fig:boxselect} Panels (a) and (b) show the expected
number of events and the sensitivity of the analysis respectively as the energy
threshold of the {\it hep} signal box is varied. The hep flux in panel (a) 
is normalized to the solar model prediction. The upper
limit of the signal box is fixed at 20~MeV.}  
\end{center}
\end{figure}

\begin{figure}[tbp]
\begin{center}
\includegraphics[width=3.3in]{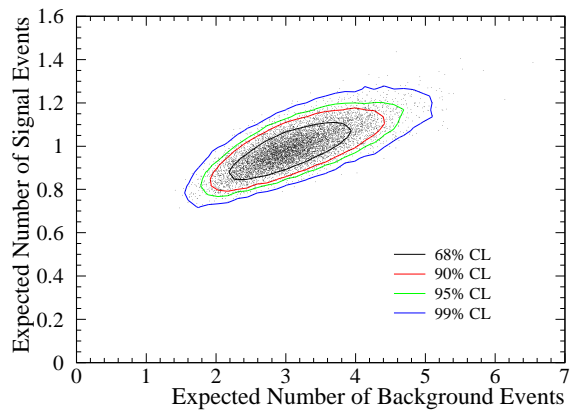}
\caption{\label{fig:pab_hep} The joint probability distribution of
signal and background events in the {\it hep} signal box assuming the
standard solar model flux of {\it hep} neutrinos.}
\end{center} 
\end{figure}

For the {\it hep} analysis the signal box was chosen to optimize the
sensitivity based on Monte Carlo simulations. The sensitivity was
defined as the mean value, from an ensemble of Monte Carlo
experiments, of the $90\%$ CL upper limit for the {\it hep} flux,
integrated over all energies using the {\it hep} neutrino spectrum
and accounting for neutrino oscillations as discussed in
Section~\ref{sec:backgrounds}, assuming the standard solar
model. Figure~\ref{fig:boxselect}a shows the predicted numbers of
signal and background events with their $1\sigma$ uncertainties as the
lower threshold of the {\it hep} signal box is changed, and
Figure~\ref{fig:boxselect}b shows the sensitivity of the analysis as a
function of the signal box threshold. There is a region between
12.5~MeV and 14.3~MeV where the sensitivity is nearly flat. Within
this range the choice of the best signal box is a compromise between
the signal to background ratio and signal acceptance. The energy range
14.3~MeV $<\mathrm{T_{\mathrm{eff}}}<$ 20~MeV was selected. In this
range, the variations of predicted signal and background levels due to
systematic uncertainties are strongly correlated, as can be seen in
Figure~\ref{fig:pab_hep}. In this signal box, $3.13\pm0.60$ background
events and $0.99\pm0.09$ signal events are expected. The contributions
to the signal and background uncertainties from the dominant sources
of systematic uncertainties are shown in Table~\ref{tab:hep_sys}.


\begin{table}[tbp]
\tablewidth{7.7cm}
\begin{center}
\caption{\label{tab:hep_sys} The effect of systematics on {\it hep} signal and background}
\begin{tabular}{ l l l } 
\tableline 
\tableline
Systematic Source & $\Delta$Bkg$(\%)$ & $\Delta$Sig$(\%)$\\ 
\tableline 
Energy scale & 13.7 & 7.6 \\
Energy resolution & 9.7 & 0.7 \\ 
Vertex accuracy & 0.3 & 2.9 \\ 
$^8$B $\nu_e$ spectrum & 0.8 & 0.0 \\ 
$\nu_{atm}$ flux & 0.3 & 0.0 \\ 
$\Delta m^2_{12}$ & 0.6 & 0.5\\ 
$\tan^2(\theta_{12})$ & 0.7 & 3.2\\ 
Cross sections & \\
~~~~CC deuteron & 0.0 & 1.1 \\ 
~~~~$\nu_{atm}$ CC QE& 0.3 & 0.0\\ 
15.1 MeV $\gamma$-rays & 0.8 & 0.0 \\
Low energy fit statistics & 3.1 & 0.0 \\
\tableline 
Combined Width & 19.1 & 9.0 \\ 
\tableline
\end{tabular} 
\end{center}
\tablecomments{This table shows the one standard deviation contributions to 
the width of the signal ($\Delta$Sig) and background ($\Delta$Bkg) 
probability distributions in the {\it hep} signal box. The combined widths 
are greater than the quadrature sums of the systematics due to correlations 
and non-linearities.}
\end{table}

\begin{figure}[h]
\begin{center}
\includegraphics[width=3.3in]{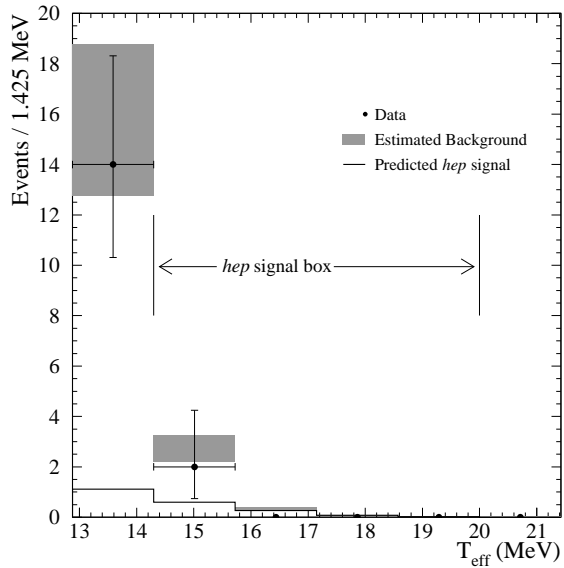}
\caption{\label{fig:hep_events}
The distribution of events in the region of the $^8$B endpoint. There
are two events in the {\it hep} signal box 14.3~MeV $<\mathrm{T_{\mathrm{eff}}}<$ 20~MeV. 
Also shown are the
estimated number of background events, including the systematic uncertainty, and the standard solar model prediction for the {\it hep} signal.}
\end{center}
\end{figure}

Two events are observed in the {\it hep} signal box. After accounting
for the effect of neutrino oscillations, this results in an upper
limit on the integral total {\it hep} neutrino flux of
$2.3\times10^4$~cm$^{-2}$s$^{-1}$ at the 90$\%$ CL. This is 2.9 times the
prediction of the standard solar model. Using a Bayesian technique
rather than the modified Feldman-Cousins approach, a limit of
$2.9\times10^4$~cm$^{-2}$s$^{-1}$ is found at the 90$\%$ CL.
The spectrum of events in the region of the signal box is
shown in Figure~\ref{fig:hep_events}. The shape agrees with the
background prediction at the $77.8\%$ CL based on Monte Carlo
simulations using a KS statistic.

This result is model-dependent, as neutrino oscillations were assumed
in the predictions of signal and background. A model-independent
search, in which a limit is placed on the $\nu_{e}$ component of the
integral {\it hep} flux, was also carried out. This search is independent of
any model of neutrino flavor change. A further event selection of
$\cos\theta_{\odot}<0.8$ was applied to remove ES events (which are
directed away from the Sun) and leave events due only to CC ($\nu_e$)
interactions. The energy threshold was selected so that the mean
background expectation was less than 0.25 events, and flux limits were
set conservatively by assuming there is no background and quoting only
the upper bound. With these criteria, a signal box of 16~MeV
$<\mathrm{T_{\mathrm{eff}}}<$ 20~MeV was selected. Without neutrino
oscillations $0.66\pm0.08$ {\it hep} events are expected in this
box. As no events are observed a limit on the $\nu_{e}$ component of
the integral {\it hep} neutrino flux of $3.1\times10^4$~cm$^{-2}$s$^{-1}$ at
the 90$\%$ CL is inferred.

Signal box selection for the DSNB search follows the approach of the
{\it hep} neutrino search. A signal box of 21~MeV $<\mathrm{T_{\mathrm{eff}}}<$
35~MeV was selected in this case. This choice was bounded at the high
end by the prior choice of the hidden energy interval and at the low
end by a desire to minimize any contribution from {\it hep}
neutrinos. In this signal box, $0.18\pm0.04$ background events are
expected. The expected number of signal events depends upon the
assumed DSNB model, but is uncorrelated with the background prediction
as different systematic uncertainties are dominant for signal and
background. The estimated values of the systematic uncertainties are
shown in Table~\ref{tab:dsnb_sys}. No events are observed in the
signal box, resulting in an upper limit of 2.3 events due to DSNB
neutrinos at the $90\%$ CL.

\begin{table}[h]
\tablewidth{7.7cm}
\begin{center}
\caption{\label{tab:dsnb_sys} The effect of systematics on DSNB signal and background}
\begin{tabular}{ l | l | l }
\tableline
\tableline
Systematic Source &  $\Delta$Bkg$(\%)$ & $\Delta$Sig$(\%)$\\
\tableline
Energy scale &  3.4 & 0.9 \\
Energy resolution & 1.7 & 0.4 \\
Vertex accuracy & 2.9 & 2.9\\
Event selection & 0.9 & 1.7\\  
$\nu$ fluxes & \\
~~~~$\nu_{atm}$ $E_{\nu}>$ 100~MeV&  8.6 &  0.0 \\
~~~~$\nu_{atm}$ $E_{\nu}<$ 100~MeV&  3.5 &  0.0\\
~~~~{\it hep} &  0.2 &  0.0 \\
Cross sections & \\
~~~~$\nu_{atm}$ CC QE&   19.2 &  0.0 \\
~~~~$\nu_{atm}$ other&  2.6 &  0.0 \\
~~~~CC deuteron &  0.2 & 1.2 \\
$\nu_{atm}$ n-multiplicity  &  0.7 &  0.0 \\
Combined Width & 24.7 & 4.3 \\
\tableline
\end{tabular}
\end{center}
\tablecomments{This table shows the one standard deviation contributions to the width of the signal ($\Delta$Sig) and background ($\Delta$Bkg) distributions in the  DSNB signal boxes. For the DSNB signal the $T=6$~MeV model from~\cite{beacom-dsnb} is used. As in the prediction for the signal and background in the {\it hep} signal box, combined widths are greater than the quadrature sums of the systematics.}
\end{table} 

To obtain a DSNB flux limit, a spectral model for the DSNB neutrinos
is required. In this paper, three models from~\cite{beacom-dsnb} and
two models from~\cite{ando-dsnb} for differential flux predictions
have been used. Table~\ref{tab:dsnb_res} shows the integral flux
predictions for these models and the 90$\%$ CL upper limits inferred
from data.

\begin{table*}[b]
\begin{center}
\caption{\label{tab:dsnb_res} DSNB flux predictions and limits}
\begin{tabular}{ l l l l l }
\tableline
\tableline
         &  \multicolumn{2}{l}{Integral Flux} & \multicolumn{2}{l}{Flux 22.9~MeV~$<E_{\nu}<$~36.9~MeV} \\
         &  \multicolumn{2}{l}{(cm$^{-2}$s$^{-1}$)} & \multicolumn{2}{l}{(cm$^{-2}$s$^{-1}$)} \\
Model    & Prediction & Upper Limit & Prediction & Upper Limit \\
\tableline
B\&S : $T = 4$~MeV &  21.1 & $1.1\times10^{4}$ & 0.19 & 93 \\
B\&S : $T = 6$~MeV &  14.1 & $1.5\times10^{3}$  & 0.66 & 72\\
B\&S : $T = 8$~MeV &  10.5 & $6.0\times10^{2}$  & 1.08 & 61\\
A\&S : NOR-L       &  28.5 & $1.3\times10^{3}$  & 1.49 & 69\\
A\&S : NOR-S-INV   &  34.9 & $2.3\times10^{3}$  & 1.06 & 70\\
\tableline
\end{tabular}
\end{center}
\tablecomments{This table shows the $90\%$ CL upper limits on the $\nu_e$ component of the DSNB flux and model predictions for different models from \cite{beacom-dsnb} (B\&S) and \cite{ando-dsnb} (A\&S). }
\end{table*}

Using these results, a limit can also be derived on the DSNB $\nu_e$
flux for neutrinos that produce electrons with kinetic energies inside
the DSNB signal box. Although the integral flux upper limits are
significantly different for these models, since their spectral shapes
are similar in the signal box, the resulting upper limits for the
neutrinos in this region vary little (see
Table~\ref{tab:dsnb_res}). Taking the median result a limit on the
DSNB $\nu_e$ flux of $70$~cm$^{-2}$s$^{-1}$ at the $90\%$ CL for
22.9~MeV $<E_{\nu}<$ 36.9~MeV is inferred. These limits and the
background prediction are in good agreement with those predicted by
\cite{beacom-dsnb}, after accounting for the difference in exposure
between their prediction and the data used in this search.

%%%%%%%%%%%%%%%%%%%%%%%%%%%%%%%%%%%%%%%%%%%%%%%%%%%%%%%%%%%%%%%%%%%%%%%%%%
%                           Conclusions                                  %
%%%%%%%%%%%%%%%%%%%%%%%%%%%%%%%%%%%%%%%%%%%%%%%%%%%%%%%%%%%%%%%%%%%%%%%%%%

\section{Conclusions}
\label{sec:conclusions}

Data from the first operational phase of SNO, with an exposure of 0.65
kilotonne-years, have been used to search for neutrinos from the {\it
hep} reaction in the Sun. No evidence for these neutrinos was
observed, and an upper limit on the integral total flux of {\it hep}
neutrinos of $2.3\times10^4$~cm$^{-2}$s$^{-1}$ has been derived at the
$90\%$ CL.  This measurement improves the previous best limit on the
{\it hep} neutrino flux, measured with the Super-Kamiokande detector,
by a factor of 6.5, but is not inconsistent with the standard solar
model. A model-independent limit on the integral {\it hep} $\nu_e$
flux, with no assumptions about neutrino oscillations, is set at
$3.1\times10^4$~cm$^{-2}$s$^{-1}$. A search for the $\nu_e$ component
of the diffuse supernova neutrino background has also been made using
SNO data. Again, no evidence for these neutrinos was found, and an
upper limit at $90\%$ CL on the $\nu_e$ component of the DSNB flux of
$70$~cm$^{-2}$s$^{-1}$ for 22.9~MeV $<E_{\nu}<$ 36.9~MeV is
inferred. This is an improvement of two orders of magnitude on the
previous $\nu_e$ limit (\cite{mb-dsnb}). The exposure of the final SNO
data set for these analyses combined across all phases of the
experiment, is expected to be approximately four times that of the
data used in this analysis.  A future search for {\it hep} and DSNB
fluxes using these data will be carried out, which is expected to
further improve upon the limits presented in this paper.

%%%%%%%%%%%%%%%%%%%%%%%%%%%%%%%%%%%%%%%%%%%%%%%%%%%%%%%%%%%%%%%%%%%%%%%%%
%                            Acknowledgments                            %
%%%%%%%%%%%%%%%%%%%%%%%%%%%%%%%%%%%%%%%%%%%%%%%%%%%%%%%%%%%%%%%%%%%%%%%%%

%\section*{ACKNOWLEDGMENTS} 
\acknowledgements
This research was supported by: Canada:
Natural Sciences and Engineering Research Council, Industry Canada,
National Research Council, Northern Ontario Heritage Fund, Atomic
Energy of Canada, Ltd., Ontario Power Generation, High Performance
Computing Virtual Laboratory, Canada Foundation for Innovation; US:
Department of Energy, National Energy Research Scientific Computing
Center, Alfred P. Sloan Foundation; UK: Particle Physics and Astronomy
Research Council. We thank the SNO technical staff for their strong
contributions.  We thank Inco, Ltd. for hosting this project.

%%%%%%%%%%%%%%%%%%%%%%%%%%%%%%%%%%%%%%%%%%%%%%%%%%%%%%%%%%%%%%%%%%%%%%%%
%                              Bibliography                            %
%%%%%%%%%%%%%%%%%%%%%%%%%%%%%%%%%%%%%%%%%%%%%%%%%%%%%%%%%%%%%%%%%%%%%%%%


\begin{thebibliography}{100}
\bibitem[Ando \& Sato(2003)]{ando-dsnb} Ando~S. \& Sato~K. 2003, Phys. Lett. B {\bf 559}, 113. Flux predictions from this paper have been increased by a factor of three, as recommended by the authors, to account for updated star formation rate data. Estimates for the $\nu_e$ DSNB fluxes were provided by the authors upon request.
\end{thebibliography}
\end{document}